\documentclass{revtex4}
\usepackage{graphicx}
\usepackage{amsmath}
\usepackage{amsfonts}
\usepackage{amssymb}

\newcommand{\ii}{\'\i}
\newcommand{\ieq}{\begin{equation}}
\newcommand{\feq}{\end{equation}}
\newcommand{\iq}{\begin{eqnarray*}}
\newcommand{\fq}{\end{eqnarray*}}
\newcommand{\iqn}{\begin{eqnarray}}
\newcommand{\fqn}{\end{eqnarray}}

\begin{document}

\begin{abstract}

The results from a series of experiments are presented whose
purpose is to explore different schemes which may lead to the
formation of pure metal plasmas in a capillary discharge with
parameters appropriate for X ray lasing. The experiments were
performed in ceramic wall capillary discharges at currents of up
to 120 kA, with an available ID of between 3 and 8 mm and with
lengths from 60 to 100 mm. Initial plasma conditions in the
capillary exploit transient hollow cathode effects in a
preionizing discharge. A laser focussed onto the back surface of
the cathode initiates both beam activity in the capillary volume
and plasma injection. To promote metal ablation into the pinch
channel of elements other than the ceramic wall material, a number
of graded ring schemes have been tried. The plasma is observed
axially using both time and energy resolved soft X-ray pin hole
images as well as from time resolved soft X-ray spectra. By
varying the rate of rise of the current of the main discharge, and
the preionizing conditions the diameter and the stability of the
Z-pinch column are seen to be affected. The ratio of the species
from the ablated wall material to the plasma formed from the
graded ring structure is found to depend both on the capillary
dimensions as well as the other operating conditions.

\end{abstract}

\title{Transient Hollow Cathode Effects and Z pinch Formation in a High Current Capillary Discharge
with a Metal Plasma}

\author{Edmund S. Wyndham}
\author{Mario Favre}
\author{Ra\'ul Aliaga Rossel}
\author{Hern\'an Chuaqui}
\author{Ian Mitchell}
\author{Jorge S. D\'\i az}
\author{Alvaro Pernas}

\affiliation{Facultad de F\ii sica, Pontificia Universidad Cat\'olica de Chile, Casilla 306, Santiago 22, Chile.}%
\date{July 2004}


\maketitle

\section*{INTRODUCTION}

The capillary discharge has lately been the object of much
experimental research with the observation of lasing in a Ne-like
Argon plasma [1]. Even discharges using stored driver energies of
$\sim$0.1J have been shown to be very bright sources of nanosecond
soft X-ray radiation in submillimeter diameter plasmas. The
importance of well collimated transient electron beams generated
by using a hollow cathode, HC, geometry has shown to be of key
importance [2] in the pulsed or THCD (transient HC discharge)
mode. The variety of high density and hot plasmas has so far been
limited to ablating wall material or using an initial filling gas.
The purpose of the present work is to attempt the formation of
metal plasmas in capillaries of up to 10 cm in length. Electron
beams emanating from the HC are essential to produce the initial
pinch conditions in the capillary. In a previous work [3] the
properties of a plasma generated in a 6 cm smooth bore alumina
capillary were presented. Using the same generator and
incorporating a series of metal rings along the length of the
tube, a significant fraction of metal plasma has been produced in
discharges from 6 to 11 cm in length and ¡n several geometries of
tube and ring structure. In all cases no initial filling gas is
present.

\section*{DESCRIPTION OF THE EXPERIMENT}

The capillary is mounted as the load of a small switched line
coaxial pulsed power generator, GEPOPU, which allows pulses to 150
kA in a nominal 120 ns pulse at 3$\cdot$ $10^{12}$ A/s. If the
line switch is shorted the capillary load may switch the current
and this mode allows a slower value of the current rise, $dI/dt$,
of $1,5\cdot10^{12}$ A/s. A small peaking gap between the line and
the load allows the application of a preionizing voltage of up to
20kV via a resistor to allow a preionizing current of $\sim$100A.
A pulsed Nd:YAG laser focussed onto a Ti bar in the hollow cathode
volume generates a plasma which injects e-beams into the capillary
volume, whose potential difference falls immediately to $\sim$800
V. The pulsed power pulse is applied between 10 and 80 $\mu$s
later. In addition to the usual voltage and current monitors, PIN
diodes, a combined XRD and Faraday cup, time resolved energy
resolved pinhole images and soft X-ray spectra in the range of 30
to 300 A using a Rowland Circle grazing incidence spectrometer.

\begin{figure}[ht]
\includegraphics[width=0.9\textwidth]{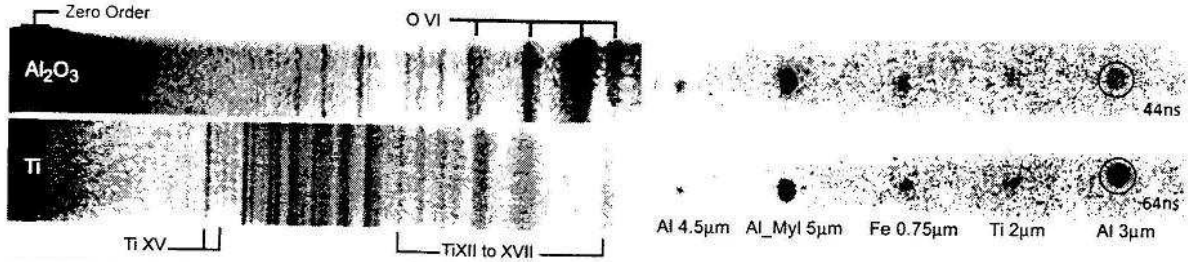}
\caption{Left. Upper: spectrum obtained in 5mm ID simple Al tube
compared with , lower, a discharge using Ti rings with a 2.8 mm
ID. Right. Two sets of 5 ns exposure filtered soft X-ray pinhole
images taken at 44 and 64 ns into the current pulse. The outline
of the available diameter for the plasma is shown as a circle.}
\end{figure}

\subsection*{Experiments Performed and Results}

An initial experiment was performed using a   6 cm alumina tube
with a 5 mm ID in which eight Ti rings of 2.8 mm ID which are
placed equidistantly. In Fig. 1 we compare the pinhole images and
the spectra between the discharge in capillaries with and without
rings. The gross temporal X-ray emission and current trace remain
unchanged, but substantial differences are seen in the spectra. It
is immediately obvious that the prominent O VI lines in the smooth
bore alumina capillary are entirely absent in the Ti ring
capillary. A number of; additional lines are visible in the Ti
ring spectrum, and many of these may be ascribed to Ti lines.
Those not assigned to Ti are other stages of Al. Limitations of
the resolution of the spectrometer do not allow a definitive
assignment of a considerable number of lines as many TiXII lines
are very close to Al VI to Al VIII lines. A series of filtered
pinholes are shown at two instants during the increasing current
period of the discharge in the Ti ring geometry. The images
indicate that the shorter wavelength radiation comes from a
central core of the Z pinch.   Some instabilities are evident from
the Ti image, but they are less marked than in longer discharges
or those of larger ID.    Further to this geometry, discharges
were performed in an 8 mm ID alumina tube with 3.8 mm ID Ti rings.
These discharges are notably less stable.

In Fig. 2 we present current, e-beam, soft X-ray activity and
pinhole images taken in an 11 cm capillary discharge. Eight Ti
rings are placed equidistantly allowing a plasma channel of 2.9
mm. For these longer discharges the capillary self switches after
about 50 ns of e-beam activity. The e-beam activity starts with a
burst of irregular and decaying amplitude and terminates with a
larger pulse which ends when the discharge current has attained
about 10 kA. During this period of build up of activity in the
hollow cathode, the tube holds the full 150kV of the applied line
voltage with a current of order 1 kA. The current rise time is
considerably slower than would be expected with the increased load
inductance of the capillary. Simulation of the load current using
a SPICE code indicates the dominant effect is that of an
increasing conductivity of the plasma during the 80 ns following
the initiation of the main current conduction. A further condition
for operation is that plasma must penetrate from the laser spark
into the capillary volume and this requires at least 20$\mu$s. The
pinhole images are not as tight as the 6 cm discharge and their
intensity is reduced.

\begin{figure}[ht]
\includegraphics[width=0.7\textwidth]{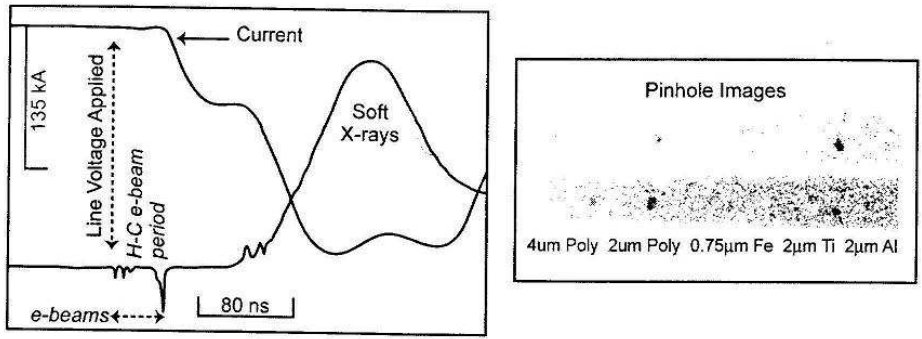}
\caption{Left. Upper: Current trace in 11 cm discharge, sowing
self switching of capillary after hollow cathode beam period has
terminated; lower, electron beams and soft X-ray emission. Right.
Two sets of 5 ns exposure filtered soft X-ray pinhole images taken
at 80 and 100 ns into the current pulse.}
\end{figure}

Figure 3 shows part of the spectrum obtained, between 35 and 150
\AA, for the long capillary discharge. The two images show the
differences in the species present when preionizing is present or
not. When not present the laser plasma expands freely into the
capillary; the external electric field is only applied when the
main line voltage is applied. In the region between 59 and 93
\AA\,
 a number of Al and Ti lines coincide, but inspite of this a number
of assignments may be attempted. The ionization stages present can
be estimated very approximately by using a CRE model such as FLY
[4]. This code, however, only models pure plasmas and Li-like and
higher stages of ionization. While Al VIII and IX stages are
present, the most intense lines, marked `(p)' of the NIST tables
are notable for their weakness. The ionization stages present as a
function of temperature and density and the expected spectrum
remain unknown for mixes of arbitrary concentrations of Al, Ti and
O. Al XI stages are barely detectable, as are Al VI. Al VII may be
clearly identified at 93.3 \AA. A group of Ti XII lines is seen at
59 \AA in the case of no preionization. It is, however, easier to
assign lines outside the range mentioned. At shorter wavelengths,
where the spectrometer rapidly loses sensitivity, a number of Ti
XIII and XV resonance lines may be seen in the second order. At
longer wavelengths these stages as well as Ti XII to XIV have a
considerable number of intense lines where only aluminium
transitions of Al IV to VI are significant, and the plasma is
substantially hotter than we would expect for emission of Al IV
and A1 V.

\begin{figure}[ht]
\includegraphics[width=0.9\textwidth]{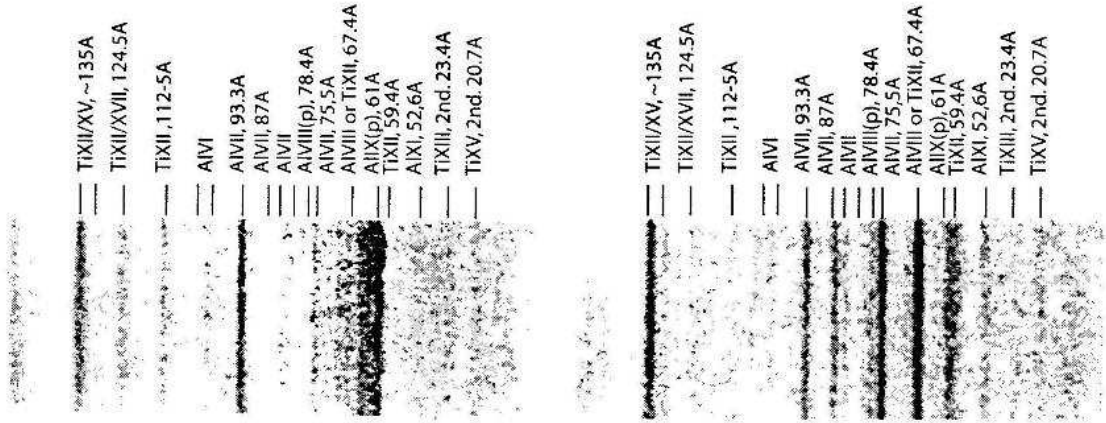}
\caption{Spectra taken in 11 cm discharge comparing the effect of
(left) prionizing current on and (right) preionizing current off.}
\end{figure}

\section*{DISCUSSION}

The plasma instabilities are less severe for the smaller of the
two tube diameters. Instabilities are consistent with a rather
wider range ionization stages in coexistance than for a single
temperature plasma. While the THCD injected plasma is sufficient
to provide a low impedance Z-pinch channel at 6 cm length, the
longer discharge is limited in X-ray emission by the significantly
lower current density. Work is required to increase the plasma
injection and beam intensity from THCD, as is theoretical modeling
to allow the precise composition of the plasma from the spectra.
Work is presently under way to produce pure Ti plasmas, in the
absence of other elements.

\section*{ACKNOWLEDGEMENTS}

We gratefully acknowledge funding from the ANDES Foundation
C-13768 and FONDECYT 1030968, both Chilean institutions, the
former private and the latter public.

\end{document}